\documentstyle [12pt]{article}

\begin{document}

\begin{flushright}
TCD--3--93 \\
March 1993
\end{flushright}

\vspace{8mm}

\begin{center}

{\Large\bf Tachyon Back Reaction \\
\vspace{3ex}
on $d = 2$ String Black Hole}  \\
\vspace{12mm}
{\large S. Kalyana Rama}

\vspace{4mm}
School of Mathematics, Trinity College, Dublin 2, Ireland. \\
email: kalyan@maths.tcd.ie   \\
\end{center}

\vspace{4mm}

\vspace{4mm}

\begin{quote}
ABSTRACT.  We describe a static solution for $d = 2$ critical
string theory including the tachyon $T$ but with its potential
$V (T)$ set to zero. This solution thus incorporates tachyon
back reaction and, when $T = 0$, reduces to the black hole solution.
When $T \neq 0$ we find that (1) the Schwarzschild horizon of the
above black hole splits into two, resembling Reissner-Nordstrom
horizons and (2) the curvature scalar develops new singularities
at the horizons. We show that these features will persist even
when $V (T)$ is nonzero. We present a time dependent extension
of our static solution and discuss some possible methods
for removing the above singularities.
\end{quote}


\newpage

\vspace{4ex}

\centerline{\bf 1. Introduction}

\vspace{4ex}

It is a challenging task to resolve the puzzles of
gravitating systems such as the nature of singularities, the end effect
of Hawking radiation, and information loss(?) inside the black hole.
Recently there has been a renewed effort
\cite{CGHS,review} to solve these problems
in the simpler context of two dimensional $(d = 2)$ systems, using the
string inspired toy models for quantum gravity, the inspiration having
come from the discovery of black holes in  $d = 2$ critical
strings \cite{W,M,rest}.

The $d = 2$ black hole for graviton-dilaton system was discovered
in an $ \rm{SL} (2, \rm{R}) / \rm{U} (1) $ gauged
Wess-Zumino-Witten model \cite{W}; as a solution
of ${\cal O} ( \alpha' ) \; \beta$-function equations \cite{CFMP,DS}
for critical
string theory with $d = 2$ target space-time \cite{M};
and in other forms \cite{rest}.
Similar toy models in $d = 2$ space-time for quantum
gravity including matter were studied first by \cite{CGHS}
and then by many others \cite{review}
with a view of solving various puzzles of gravitating systems
in this simpler context. Most of these models have mainly
studied the graviton-dilaton system.

For a more complete story, one should also include tachyon, the only
remaining low energy degree of freedom for $d = 2$ strings (and which is
a massless excitation for $d = 2$). Though important, its
inclusion results in non linear equations which are solved
only in a few asymptotic cases \cite{DL,review}.
The solution of $\beta$-function equations for
$d = 2$ strings including tachyons is not known.
However, solving for tachyon in the $d = 2$ string
black hole background leads to singular behaviour \cite{W,M}.
Thus it is important to
understand the back reaction of tachyons for $d = 2$ critical strings,
especially  because of its importance as a model for $d = 2$ quantum
gravity.

In this work we describe a static solution,
recently announced in \cite{R}, of the $\beta$-function
equations for the low energy $d = 2$ critical string theory
including tachyon $T$ but not its potential $V(T)$.
This solution incorporates tachyon back reaction.
The solution has, beside the black hole mass parameter,
a new parameter $\epsilon$ which is
a measure of tachyon strength. We find that: (1) The Schwarzschild horizon
of the previous black hole splits into two, resembling
Reissner-Nordstrom horizons. However, the solution cannot be
analytically continued in between the two ``horizons''.
(2) The curvature scalar $R$ has a point singularity, as for
$\epsilon = 0$, but when $\epsilon > 0$ develops new singularities
at the horizons. Hence this solution is not a black hole
solution in the usual sense, though for $\epsilon = 0$ it is.
Though we needed to set $V = 0$ to obtain an explicit solution, we
show later that the features of this solution will
persist even when $V$ is correctly taken into account.
Thus this shows that when tachyon back reaction is included,
the $d = 2$ string does not admit any static solution which can be
interpreted as a black hole in the usual sense. We also
present a time dependent extension of our solution which, however,
does not alter the features of the static solution, and then
discuss critically some possible methods of removing the singularities.

In section 2 we present $\beta$-function equations in various forms
and obtain a static solution of these equations in section 3
assuming that the tachyon potential vanishes. We discuss the features
of this solution in section 4 and show, in section 5, that
these features will persist even when the tachyon potential is nonzero.
In section 6 we present a time dependent extension of our static
solution and then conclude with a
critical discussion of possible methods of removing the singularities,
atleast the new ones.

\vspace{4ex}

\centerline{\bf 2. $\beta$-function Equations}

\vspace{4ex}

In this section we give $\beta$-function equations in various forms
for the low energy $d = 2$ critical string theory including
graviton $(G_{\mu \nu})$, dilaton $(\phi)$, and tachyon $(T)$ fields.
We consider mostly the ``static'' case.
The sigma model action of the $d = 2$ critical string theory for
$G_{\mu \nu}, \; \phi$, and $T$, in
a notation similar to that of \cite{M}, is given by
\[
S_{\rm sigma} = \frac{1}{8 \pi \alpha'} \int d^2 \tilde{x} \sqrt{g}
( G_{\mu \nu} \nabla x^{\mu} \nabla x^{\nu} + \alpha' R \phi
+ 2 T )
\]
where $\tilde{x}$ are the world sheet coordinates and
$ x^{\mu}, \; \mu = 0, \, 1$ are the target space coordinates.
The conformal invariance of the above action requires the following
$\beta$-function equations to be satisfied:
\begin{eqnarray}\label{beta}
R_{\mu \nu} + \nabla_{\mu} \nabla_{\nu} \phi
+ \nabla_{\mu} T \nabla_{\nu} T & = & 0 \nonumber \\
R + (\nabla \phi)^2 + 2 \nabla^2 \phi + (\nabla T)^2
+ 4 \gamma K & = & 0 \nonumber \\
\nabla^2 T + \nabla \phi \nabla T - 2 \gamma K_T & = & 0
\end{eqnarray}
where
$ \gamma = \frac{- 2}{\alpha'}, \;
K = 1 + \frac{V}{4 \gamma} , \;
V = \gamma T^2 + {\cal O} (T^3)$ is the tachyon potential
and $K_T \equiv \frac{d K}{d T}$.
These equations also follow from the target space effective action
\begin{equation}\label{target}
S = \int d^2 x \sqrt{G} \, e^{\phi} \, ( R - (\nabla \phi)^2
+ (\nabla T)^2 + 4 \gamma K ) \; .  \nonumber
\end{equation}
Note from equation (\ref{target}) that the dilaton field
$e^{- \frac{\phi}{2}}$ acts as a coupling constant.
Consider now the
target space conformal gauge $ d s^2 = e^{\sigma} d u \, d v $
where $u = x^0 + x^1$ and $v = x^0 - x^1$. In this gauge
the only nonzero connection ($\Gamma_{\mu \nu}^{\kappa}$)
and curvature ($R_{\mu \nu}$) components are
\begin{equation}
\Gamma_{u u}^u = \partial_u \sigma \; , \;\;
\Gamma_{v v}^v = \partial_v \sigma \; ; \;\;
R_{u v} = \partial_u \partial_v \sigma
\end{equation}
and equations (\ref{beta}) become
\begin{eqnarray}\label{betauv}
\partial_u^2 \phi - \partial_u \sigma \partial_u \phi
+ ( \partial_u T )^2  & = & 0 \nonumber \\
\partial_v^2 \phi - \partial_v \sigma \partial_v \phi
+ ( \partial_v T )^2 & = & 0 \nonumber \\
\partial_u \partial_v \sigma + \partial_u \partial_v \phi
+ \partial_u T \partial_v T & = & 0 \nonumber \\
\partial_u \partial_v \phi + \partial_u \phi \partial_v \phi
+ \gamma K e^{\sigma} & = & 0 \nonumber \\
2 \partial_u \partial_v T + \partial_u \phi \partial_v T
+ \partial_v \phi \partial_u T - \gamma K_T e^{\sigma} & = & 0 \; .
\end{eqnarray}
Defining new coordinates
$\xi = u v, \; \chi = u/v$,
and, for convenience $e^{\Sigma} = \gamma \xi e^{\sigma}$,
equations (\ref{betauv}) can be written as
\begin{eqnarray}\label{betaxichi}
( \phi_{1 1} - \Sigma_1  \phi_1 + T_1^2 )
+ ( \phi_{0 0} - \Sigma_0  \phi_0 + T_0^2 ) & = & 0 \nonumber \\
( \Sigma_{1 1} + \phi_{1 1} + T_1^2 )
- ( \Sigma_{0 0} + \phi_{0 0} + T_0^2 ) & = & 0 \nonumber \\
2 \phi_{0 1} - \Sigma_1 \phi_0 - \Sigma_0 \phi_1
+ 2 T_0 T_1 & = & 0 \nonumber \\
( T_{1 1} + \phi_1 T_1 - \frac{1}{2} e^{\Sigma} K_T )
- ( T_{0 0} + \phi_0 T_0 ) & = & 0 \nonumber \\
( \phi_{1 1} + \phi_1^2 + e^{\Sigma} K )
- ( \phi_{0 0} + \phi_0^2 ) & = & 0
\end{eqnarray}
where
$( \cdots )_0 = (\chi \frac{d}{d \chi}) ( \cdots )$ and
$( \cdots )_1 = (\xi \frac{d}{d \xi}) ( \cdots )$.
In the following we first specialise to the static case
where the fields depend only on $\xi$. Then
equations (\ref{betaxichi}) become
\begin{eqnarray}\label{salmon}
\Sigma_{1 1} + \phi_1 \Sigma_1 = \Sigma_{1 1} + \phi_{1 1}
+ T_1^2 & = & 0 \nonumber \\
T_{1 1} + \phi_1 T_1 - \frac{1}{2} e^{\Sigma} K_T =
\phi_{1 1} + \phi_1^2 + e^{\Sigma} K & = & 0  \; .
\end{eqnarray}

To proceed further, we first note that
the fields can be expected to evolve depending on the local metric
given by $\Sigma$ and
hence look for their solutions in terms of $X(\xi) \equiv \Sigma_1$.
Letting $F(X) = \phi_1, \;
( \cdots )' = \frac{d}{d X} ( \cdots )$, and noting that
$( \cdots )_1 = ( \cdots )' X_1$, equations (\ref{salmon}) give
\begin{eqnarray}\label{squid}
X_1 + X F = 0 , \; \; \;
T_1^2 - X F (1+ F') & = & 0 \nonumber \\
X F F'' + (X F' - F) (1 + F') + e^{\Sigma} T_1 K_T (X F)^{-1}
& = & 0  \; .
\end{eqnarray}
The curvature scalar is given by
$R = - 4 \gamma e^{- \Sigma} X F$.
Dividing the last equation above
by $X F ( 1 + F')$ and changing the
independent variable from $\xi$ to $X$ the $\beta$-function equations
for the static case can be written as
\begin{eqnarray}\label{betax}
F \Sigma' + 1  & = & 0 \nonumber \\
X \phi' + 1  & = & 0 \nonumber \\
X F T'^2 - (1 + F') & = & 0 \nonumber \\
\frac{F''}{1 + F'} + \frac{F'}{F} - \frac{1}{X}
- \frac{T' K_T e^{\Sigma}}{X F (1 + F')} & = & 0
\end{eqnarray}
where  $X$ and $\xi$ are related by $X_1 = - X F$.
Equations (\ref{betax})
can be simplified further, for example, as follows. Let
\begin{equation}\label{l}
L' = \frac{T' K_T e^{\Sigma}}{X F (1 + F')} \; .
\end{equation}
In this case the last
equation in (\ref{betax}) can be integrated to give
\begin{equation}
F (1 + F') = a^2 X e^L
\end{equation}
where $a^2 $ is a constant. This immediately gives
$T_1^2 = a^2 X^2 e^L $ from which we obtain
\begin{equation}
T' = - a F^{- 1} e^{\frac{L}{2}} \; .
\end{equation}
Substituting these expressions in (\ref{l})
and replacing $L$ by a new field $H = a e^{\frac{L}{2}}$
the $\beta$-function equations become
\begin{eqnarray}\label{betah}
F \Sigma' + 1  & = & 0  \nonumber \\
F T' + H & = & 0 \nonumber \\
1 + F' - \frac{X H}{F} & = & 0 \nonumber \\
H' + \frac{K_T e^{\Sigma}}{2 X^2 F} & = & 0
\end{eqnarray}
and
\begin{equation}\label{integral}
X F + F^2 - X^2 H^2 + K e^{\Sigma} = 0  \; .
\end{equation}
The above equation forms an integral of the set of
differential equations (\ref{betah}).

We present one more version of the $\beta$-function equations
which appears simpler but is just as hard to solve as other versions.
However, this form will be useful later.
Note that equations (\ref{betah})
are invariant under the scaling
\begin{eqnarray}
X & \rightarrow & \lambda X \nonumber \\
F & \rightarrow & \lambda F \nonumber \\
\Sigma & \rightarrow & \Sigma + 2 \ln {\lambda}
\end{eqnarray}
where $\lambda$ is a constant. Hence define a set of
scale invariant variables \cite{nonlinear}
\begin{equation}\label{xfs}
x = - X^2 e^{- \Sigma} \; , \;\;
f = \frac{F}{X} \; , \;\;
s = x \Sigma' \; .
\end{equation}
In terms $x$ and $f$ equations
(\ref{betah}) and (\ref{integral}) become
\begin{eqnarray}\label{betalie}
s f + 1 & = & 0 \nonumber \\
x (1 + 2 f) \frac{d T}{d x} + H & = & 0 \nonumber \\
x (1 + 2 f) \frac{d f}{d x} + \frac{K}{x} & = & 0 \nonumber \\
x (1 + 2 f) \frac{d H}{d x} - \frac{K_T}{2 x} & = & 0
\end{eqnarray}
and
\begin{equation}\label{xfintegral}
f + f^2 - H^2 - \frac{K}{x} = 0 \; .
\end{equation}
The curvature scalar is given by $R = 4 \gamma x f$.

We now consider the $\beta$-function equations (\ref{beta}) in
the Schwarzschild
gauge $d s^2 = - G (d x^0)^2 + G^{- 1} (d x^1)^2$.
For the static case we have $G = G(x_1)$.
Equations (\ref{beta}) then become
\begin{eqnarray}\label{betasch}
G'' + G' \phi' & = & 0 \nonumber \\
\phi'' + T'^2 & = & 0 \nonumber \\
G \phi'' + \phi' ( G' + G \phi' ) + 4 \gamma K & = & 0 \nonumber \\
G T'' + T'( G' + G \phi' ) - 2 \gamma K_T & = & 0
\end{eqnarray}
where in equations (\ref{betasch}) and (\ref{confsch})
$(\cdots)'$ denotes $\frac{d}{d x^1} (\cdots)$ and
$\gamma, \; K$, and $K_T$ are as defined before.
These equations appear to have no connection with
equations (\ref{salmon}) in the conformal gauge.
However, if we define ${\cal G}$ and $\psi$ by
\begin{equation}\label{confsch}
G = 4 \gamma e^{- {\cal G}} \; ,  \; \;
\psi' = \phi' - {\cal G}' \; ,
\end{equation}
then equations for ${\cal G}, \; \psi$, and $T$ become
exactly the same as equations (\ref{salmon}) for
$\Sigma, \; \phi$, and $T$ respectively. Thus this implies
that restricting
the fields $\Sigma, \; \phi$, and $T$ to depend only on $\xi$
corresponds to the static case. Henceforth, in this paper we will
work only in the conformal gauge.

Finally, let us briefly discuss the residual
transformations which preserve both the conformal gauge
$d s^2 = e^{\sigma} d u d v$ and the static nature
of the fields. Such transformations are given by
\begin{equation}\label{resgg}
u = \tilde{u}^b , \; \; v = \tilde{v}^b
\end{equation}
where $b$ is an arbitrary constant.
Under transformations (\ref{resgg}) we have
\begin{eqnarray}
\xi & = & \tilde{\xi}^b \nonumber \\
(\cdots)_{\tilde{1}} & = & b (\cdots)_1  \nonumber \\
\tilde{\sigma} & = & \sigma + (b - 1) \ln \tilde{\xi} + 2 \ln b \nonumber \\
\tilde{\Sigma} & = & \Sigma + 2 \ln b \nonumber \\
\tilde{X} & = & b X \nonumber \\
\tilde{F} & = & b F
\end{eqnarray}
where $(\cdots)_1$ denotes
$( \xi \frac{d}{d \xi} ) (\cdots)$  as before. Also equations
(\ref{salmon}) remain invariant under these transformations.
Thus, for example, one may scale both $X$ and $F$ by an arbitrary
constant using the residual conformal gauge transformations
(\ref{resgg}).

\vspace{4ex}

\centerline{\bf 3. Static Solutions}

\vspace{4ex}

We now consider the solutions of $\beta$-function equations described
in last section. First let $T = 0$. In this case,
the general solution of equations
(\ref{beta}) is given by
the two dimensional black hole solution presented in \cite{M} which
corresponds to the static case. This solution can be written as follows.
$T = 0$ implies that $H = K_T = 0$ in equation
(\ref{betah}). We then obtain
$F = 1 - X$ where, using  transformations (\ref{resgg}),
the integration constant is set equal to $1$.
Equation (\ref{integral})
then gives $e^{\Sigma} = X - 1$. From
$X_1 = - X F$ it follows that
$X = \alpha (\xi + \alpha)^{- 1}$
where $\alpha$ is a constant related to the black hole mass
parameter $a$ of ref. \cite{M} by $ a = - \gamma \alpha$
(see section 4 also). This
constitutes the general solution of equation (\ref{beta}) when
$T = 0$.

However, by setting $T = 0$ we loose all information about tachyon
back reaction. It is important to incorporate this back reaction
particularly in view of the hints of instability found in various works
\cite{W,M,DL} when tachyon is included. The main difficulty when
$T \neq 0$ is that the resulting equations are hard to solve explicitly.
In this section we present a static solution when $T \neq 0$ thus
including the tachyon back reaction.

Notice that equations (\ref{salmon})
can be solved explicitly if one assumes that the tachyon potential
vanishes, {\em i.e.}\ , $V (T) = 0$.
The solution obtained in this approximation
incorporates tachyon back reaction and exhibits new features which,
we show later, will persist even when $V (T) \neq 0$.

When $V = 0, \; K = 1$ and (\ref{betah}) implies that
$H = {\rm constant}$. Let $H^2 = \epsilon (1 + \epsilon)$
where $\epsilon \geq 0$ (equivalently $\epsilon$ can be
$\leq - 1$) so that $T_1^2 = X^2 H^2 \;  \geq 0 $.
We then have
$F (1 + F') = \epsilon (1 + \epsilon) X $.
Note that this equation is invariant under $X \rightarrow \lambda X$
and $F \rightarrow \lambda F$ where $\lambda$ is a constant.
This invariance suggests the substitution \cite{nonlinear2}
$X = e^s , \; F = e^s \tilde{F}$. Integrating the resulting equation
gives
\begin{equation}\label{guppie}
(F - \epsilon X)^{\epsilon} \;
(F + (1 + \epsilon) X)^{1 + \epsilon} = {\rm constant} \; .
\end{equation}
Note that for $\epsilon = 0$ (hence $T = 0$) in the above expression
we get the black hole solution of \cite{M}.

In principle, this forms the complete solution. However it is difficult
to understand this solution --- to integrate $X_1 = - X F$, to understand
how the various fields $\Sigma, \; \phi, \; T$, and the curvature
scalar $R$ behave, {\em etc.}\ . Therefore we choose a different
parametristaion for the above solution. Let
\begin{equation}\label{scallop}
F - \epsilon X = l B \tau^{-1} ; \;
F + (1 + \epsilon) X = B \tau^{\delta - 1}
\end{equation}
where $\tau \geq 0$ is a new parameter, $l = \pm 1, \;
\delta = (1 + 2 \epsilon) (1 + \epsilon)^{-1}, \; \epsilon \geq 0, \;
B = A (1 + 2 \epsilon)$ and $A$ is a constant.
Note that for $l = - 1$
the constant in equation (\ref{guppie}) need not be real and hence
our choice of $l$ constitutes minimal analytic continuation.
Equations (\ref{scallop}) give
\begin{equation}
X = A \tau^{-1} (\tau^{\delta} - l); \;
F = A \tau^{-1} (\epsilon \tau^{\delta} + l (1 + \epsilon))
\end{equation}
from which it follows that
$(1 + \epsilon) \tau  \dot{X} = F$ where
$\dot{(\cdots)} = \frac{d (\cdots)}{d \tau}$.
Equations (\ref{betah}) or (\ref{salmon}) then give
\begin{eqnarray}
\dot{\phi} & = & - X^{-1} \dot{X}  \nonumber \\
\dot{\Sigma} & = & - ((1 + \epsilon) \tau)^{-1}  \nonumber \\
\dot{T} & = & - \sqrt{\delta - 1} \tau^{-1}   \nonumber \\
\tau_1 & = & - (1 + \epsilon) \tau X
\end{eqnarray}
which can be integrated to obtain
\begin{eqnarray}\label{whale}
e^{\phi} & = & \beta_0 \tau (\tau^{\delta} - l)^{-1} \nonumber \\
T & = & - \sqrt{\delta - 1} \ln\tau   \nonumber \\
e^{\Sigma} & = & - l B^2
\tau^{- \frac{1}{(1 + \epsilon)}} \nonumber \\
\int_0^{\tau} d \tau (\tau^{\delta} - l)^{-1}
& = & A (1 + \epsilon) \ln(\frac{\alpha}{m \xi})
\end{eqnarray}
where $\beta_0$ and $\alpha$ are constants and
$m = \pm 1$ is the sign of $\xi$.
The curvature scalar $R$ is given by
\begin{equation}\label{dolphin}
R = 4 \gamma (1 + 2 \epsilon)^{-2} \tau^{- \delta}
(\tau^{\delta} - l) (\epsilon \tau^{\delta} + l (1 + \epsilon)) \; .
\end{equation}
Equations (\ref{whale}) and (\ref{dolphin}) form
the solution of equations (\ref{salmon}) when $V = 0$.
Using the transformations (\ref{resgg}) we can set $B = 1$.
If  $\epsilon = 0$ and $\tau$ is eliminated, the above expressions
reduce to those given in the beginning of this section.

For future use we also write down the expressions for
$x$ and $f$ defined in (\ref{xfs}). For $\epsilon = 0$ we have
\begin{equation}\label{xf1}
x = \frac{X^2}{1 - X},  \; \; \;
f = \frac{1 - X}{X}
\end{equation}
where $X = \alpha (\xi + \alpha)^{- 1}$.
For $\epsilon \neq 0$ we get from (\ref{whale})
\begin{eqnarray}\label{xf2}
x & = & l (1 + 2 \epsilon)^{-2} \tau^{- \delta}
(\tau^{\delta} - l)^2    \nonumber \\
f & = & \frac{\epsilon \tau^{\delta} + l (1 + \epsilon)}
{\tau^{\delta} - l}  \; .
\end{eqnarray}

\newpage

\vspace{4ex}

\centerline{\bf 4. Features of the static solutions}

\vspace{4ex}

Let us now consider in detail the static solution given by
equations (\ref{whale}) and (\ref{dolphin}). The $\tau$-integration
in (\ref{whale}) cannot be done in a closed invertible form  except when
$\epsilon = 0 \; {\rm or} \; \infty$. However to understand the solution
and its salient features it is not necessary to do this integration.
The integrand $(\tau^{\delta} - l)^{- 1}$ is regular for
$\epsilon \geq 0$ and is well defined for $\tau \geq 0$. We only need
from the equation relating $\tau$ and $\xi$ the value of $\xi$ for
a given $\tau$. Since the integrand varies monotonically with
respect to $\epsilon$, this information is easy to obtain
qualitatively for generic $\epsilon$.

\vspace{1ex}

\begin{flushleft}
\underline{\bf $\epsilon = 0$ }
\end{flushleft}

\vspace{1ex}

\noindent This gives the solution obtained in \cite{M}.
In this case we get from (\ref{whale})
\begin{equation}
\ln (\tau - l) = \ln ( \frac{\alpha}{m \xi} )
\end{equation}
where $\tau \geq 0, \; \alpha$ is a constant
and we have set $B \; ( = A ) = 1$.
Using the expression for
$e^{\Sigma}$ in (\ref{whale}) and its definition
$e^{\Sigma} \equiv \gamma \xi e^{\sigma}$ where
$\gamma = - \frac{2}{\alpha'}$
we see that $l = {\rm sign} (\xi)$ so that
$e^{\sigma} \geq 0$. We first take $l = m = 1$ and write
\begin{equation}
\tau - 1 = \frac{\alpha}{\xi} \; .
\end{equation}
For $\epsilon = 0$ the above equation can be
considered to be valid for $- \infty \leq \tau \leq \infty$.
We then obtain, with $\beta_0$ a constant,
\begin{eqnarray}\label{krill}
e^{\phi} & = & \frac{\beta_0}{\alpha}
(\xi + \alpha) \nonumber \\
e^{- \sigma} & = & - \gamma (\xi + \alpha)
\end{eqnarray}
which is the solution of \cite{M} if we set
$a = - \gamma \alpha = \beta_0$ where
$a$ is the black hole mass parameter
of \cite{M}. This solution
is asymptotic to flat space time at $\xi = \infty$,
has a horizon at $\xi = 0$, and a singularity
at $\xi = - \alpha$ where the curvature scalar $R$
becomes divergent. For more details see \cite{M}.

However, in obtaining the above solution we let $\tau$
take negative values also. This is alright for $\epsilon = 0$.
But when $\epsilon > 0, \; \; \tau^{\delta}$ in the integrand
is not well defined for $\tau < 0$. Therefore we always restrict
$\tau$ to be $\geq 0$. We then proceed as
follows. Let $\xi \geq 0$. Then $l = m = 1$ so that the metric is
well defined as explained above. The range $\infty \geq \xi \geq 0$
outside the horizon
then corresponds to $1 \leq \tau \leq \infty$. We call this Branch I.
When $\xi \leq 0$ we have $l = m = - 1$ and the
range $- \alpha \geq \xi \geq 0$ inside the horizon
corresponds to $0 \leq \tau \leq \infty$. We call this Branch II.
These two branches describe the solution given above
when $\epsilon = 0$. We assume that these are also the
relevent branches when $\epsilon > 0$.

We now make a few comments on the mass of the black
hole,  $M_{bh}$, which can be
calculated using the behaviour of the metric
in the asymptotically flat region following the
methods of Witten \cite{W} or those of, for example, \cite{bhmass}.
We find it convenient to use the formula given by T. Tada and
S. Uehara in \cite{bhmass} which, in our notation, reads as
\begin{equation}\label{bhm}
M_{bh} =  (- \gamma)^{\frac{1}{2}}
e^{\phi} (e^{- \Sigma} \phi_1^2 + 1)
\end{equation}
where $(\cdots)_1 = (\xi \frac{d}{d \xi}) (\cdots)$ as before.
Using (\ref{krill}) we obtain
\begin{equation}
M_{bh} =  (- \gamma)^{\frac{1}{2}} \beta_0 \; .
\end{equation}
Note further that $M_{bh}$ as
given in (\ref{bhm}) is conserved, {\em i.e.}\ ,
\begin{equation}
M_1 = 0
\end{equation}
as can be easily verified.

\vspace{1ex}

\begin{flushleft}
(ii) \underline{\bf $\epsilon > 0$}
\end{flushleft}

\vspace{1ex}

\noindent This case includes the tachyon
back reaction. Consider branch I, {\em i.e.}\ ,
$\tau \geq 1$ and $l = m = 1$. The integral
$\int_0^{\tau} d \tau (\tau^{\delta} - 1)^{- 1}$
tends to $- \infty$ as $\tau \rightarrow 1_+$  and tends to a finite
positive semidefinite value ($ = A (1 + \epsilon)
\ln \frac{\alpha}{\xi_+}$, by definition)
as $\tau \rightarrow \infty$. Thus we have
$\infty \geq \xi \geq \xi_+ > 0$ as $1 \leq \tau \leq \infty$.
Similarly for branch II, the integral
$\int_0^{\tau} d \tau (\tau^{\delta} + 1)^{- 1}$
tends to a finite
positive semidefinite value ($ = A (1 + \epsilon)
\ln \frac{\alpha}{\xi_-}$, by definition)
as $\tau \rightarrow \infty$. It then follows
that $- \alpha \leq \xi \leq - \xi_- < 0$ as
$0 \leq \tau \leq \infty$. Moreover
$\xi_{\pm} \rightarrow 0$ as $\epsilon \rightarrow 0$. This implies that
the horizon which was located at $\xi = 0$ when $\epsilon = 0$
now splits into two located at $\xi = \pm \xi_{\pm}$. Thus the horizon
resembles Schwarzschild horizon for $\epsilon = 0$ and
Reissner-Nordstrom horizon for $\epsilon > 0$.

That $\tau = \infty$ does indeed correspond to the location of
horizon can also be seen by the
zeroes of the metric $G_{\mu \nu}$ in Schwarzschild coordinates
$(r, t)$ defined by
$d \xi = 2 (- \gamma)^{\frac{1}{2}} \xi e^{- \Sigma} d r$ and
$d \chi = 2 (- \gamma)^{\frac{1}{2}} \chi d t$. The metric is given by
\begin{equation}\label{rtmetric}
d s^2 = e^{\Sigma} d t^2 - e^{- \Sigma} d r^2
\end{equation}
where $e^{\Sigma} = \gamma \xi e^{\sigma}$ and
$\gamma = - \frac{2}{\alpha'}$. Equation (\ref{whale}) implies that
$e^{\Sigma} \rightarrow 0$ as $\tau \rightarrow \infty$ for
any $\epsilon \geq 0$. Hence $\tau = \infty$
corresponds to a horizon.
Note also that in Schwarzschild coordinates
the fields are indeed static, {\em i.e}\ , independent of
time coordinate $t$ and that the asymptotically flat region is given
by $r \rightarrow \infty$.

Though  for $\epsilon > 0$
the metric components $G_{\mu \nu}$ have the
right zeroes and poles at $\tau \rightarrow \infty$, it is not
really a horizon in the usual sense because
$G_{\mu \nu}$ cannot be made regular by transforming coordinates.
Thus $\tau = \infty$ is not just a coordinate singularity.
This can be seen by evaluating the curvature scalar $R$ at
$\tau = \infty$. We see from equation (\ref{dolphin}) that
$R \rightarrow \infty$ as $\tau \rightarrow \infty$ with a strength
proportional to $\epsilon$ for small values of $\epsilon$. Thus these
new horizons have curvature singularities. Also,
unlike in the case of Reissner-Nordstrom black hole, the above solutions
cannot be extended analytically into the region
$- \xi_- \leq \xi \leq \xi_+$, retaining
their correct limiting behaviour
as $\epsilon \rightarrow 0$. Hence our solution is not
really a black hole solution, though for $\epsilon = 0$ it is.

The curvature scalar $R$ has another singularity in branch II
at $\tau = 0$ as can be seen from equation (\ref{dolphin}). This is
a singularity inside the inner horizon and is present even when
$\epsilon = 0$, {\em i.e.}\ , $T = 0$. This is the singularity
present in the blackhole solutions of \cite{W,M}.

Note further that at $\tau = \infty$ the dilaton field
$e^{\phi} \rightarrow \tau^{- \frac{\epsilon}{1 + \epsilon}}
\; ( \rightarrow 0)$. Since $e^{- \frac{\phi}{2}}$ acts as a coupling
constant in equation (\ref{target}), it follows that one
is dealing with a strong coupling regime near the horizon when
$\epsilon > 0$, {\em i.e}\ , when tachyon back reaction is included.

Consider now the ``black hole'' mass $M_{bh}$ calculated from
the behaviour of the fields in our solution
in the asymptotically flat region.
It can be easily seen that in the asymptotically flat region
(branch I, $\tau \rightarrow 1_+$) the tachyon field is negligible.
(This is true even when the tachyon potential $V \neq 0$.
See section 5 and, for example, \cite{DL}).
Therefore, equation (\ref{bhm}) can still be applied
with negligible error to calculate
$M_{bh}$ from the fields in the asymptotically flat region. We obtain
\begin{equation}
M_{bh} = \frac{(- \gamma)^{\frac{1}{2}} \beta_0}
{1 + 2 \epsilon}  \; .
\end{equation}
Thus the behaviour of the fields in the asymptotically flat
region is not giving any
indication of the new singularities that arise as a result of
the tachyon back reaction. However, the formula (\ref{bhm}) cannot
be relied upon since it is not conserved when one uses the equations
of motion (\ref{salmon}) which include tachyon also. That is,
\begin{equation}
( M_{bh} )_1 \neq 0 \; .
\end{equation}
It is not clear if any conserved quantity exists when tachyons are
included. We could not find any, except the trivial ones given
in the equations of motion (\ref{salmon}) or their linear combinations.

\vspace{4ex}

\centerline{\bf 5. Effect of tachyon potential, $V (T)$}

\vspace{4ex}

In this section we consider the tachyon potential, $V$. It was
necessary to assume that $V = 0$ in deriving our solution.
We show below that even when $V \neq 0$ the features we
found in our solution will persist.

It is convenient to view
the tachyon equation in (\ref{salmon})
as an equation for  an (anti)damped oscillator where
the  couplings to graviton and dilaton provide the damping force and
the tachyon potential $V$ provides the restoring force.
This would mean that as one moves in from the asymptotically flat region
(branch I, $\tau \rightarrow 1_+$) towards the horizon
($\tau \rightarrow \infty$), the tachyon will interact with
gravitational field acquiring a large kinetic energy. Then
the tachyon potential can be neglected in comparison to its kinetic
energy. That this is actually what happens can be seen
by calculating the kinetic
$( (\nabla T)^2 )$ and potential $( V )$ energy terms in the
action (\ref{target})
(or, equivalently, the damping $( \phi_1 T_1 )$
and the potential $( e^{\Sigma} K_T )$ terms in the tachyon equation
in (\ref{salmon}) ). Using the expressions
$ (\nabla T)^2 = 4 \gamma e^{- \sigma} T_1^2 , \,
T = - \sqrt{\delta -1} \ln\tau , \;
\phi_1 T_1 = \sqrt{\epsilon (1 + \epsilon)} X F$, and
$e^{\Sigma} K_T = e^{\Sigma} (\frac{T}{2} + {\cal O}(T^2)) $
and the solution given in equation (\ref{whale}) one sees that
tachyon kinetic energy indeed dominates its potential energy
away from the asymptotically flat region.
Hence, neglecting tachyon potential is a valid approximation far
inside from the asymptotically flat region (branch I, $\tau >> 1$).

Asymptotically, however, the kinetic and potential enrgy of the
tachyon are of the same order and hence $V$ cannot be
neglected\footnote{Indeed, in the asymptotically flat
region $\tau \rightarrow 1_+$,
$T_V = \sqrt{T}$ where $T_V$ is the correct tachyon
solution including its potential $V (T) = \gamma T^2$. This relation
can be worked out easily by substituting the expressions for
graviton and dilaton that follow from equation (\ref{whale}) into
the tachyon equation given in (\ref{salmon}) and solving for
tachyon near $\tau \rightarrow 1_+$ as we will show. Or, it can
also be seen easily from a calculation in ``linear dilaton vacuum''
similar to that of de Alwis and Lykken in ref. \cite{DL}.}.
But it is reasonable to expect that the correct
asymptotically flat
solution when $V \neq 0$ can be matched at some point to
(\ref{whale}) which becomes more and more valid as one nears the
horizon. That this is likely to be the case can be seen
by performing one iterative calculation when $V \neq 0$ as follows.
Near the asymptotically flat region, let $\tau = 1 + y$ where
$y > 0$ is small. Denoting
$\dot{(\cdots)} = \frac{d}{d \tau} (\cdots)
= \frac{d}{d y} (\cdots) $,
the dilaton and tachyon equations in (\ref{salmon}) become
\begin{eqnarray}\label{mussel}
\ddot{\phi} + \tau^{- 1} \dot{\phi} + (1 + \frac{T^2}{4})
\frac{e^{\Sigma}}{\tau_1^2} & = & 0    \nonumber \\
\ddot{T} + \tau^{- 1} \dot{T}
- \frac{T e^{\Sigma}}{4 \tau_1^2} & = & 0
\end{eqnarray}
where we have taken $V (T) = \gamma T^2$. Note that
$\tau^{- 1} \dot{T}$ and $- \frac{T e^{\Sigma}}{4 \tau_1^2}$
act like effective damping and restoring forces  respectively in the
tachyon equation. For $y \rightarrow 0$ the solutions
(\ref{whale}) become, with $B = 1$,
\begin{eqnarray}
\tau_1 & = &  - y + \cdots \nonumber \\
e^{\Sigma} & = & - 1 + \frac{y}{1 + \epsilon} + \cdots \nonumber \\
\phi & = & - \ln y + {\rm constant}
+ \frac{2 + \epsilon}{2 (1 + \epsilon)} \; y + \cdots
\end{eqnarray}
where $\cdots$ denote higher order terms in $y$.
The restoring force is given by
\begin{equation}
- \frac{T e^{\Sigma}}{4 \tau_1^2} = \frac{T}{4 y^2}
(1 - y + \cdots)  \; .
\end{equation}
It follows from the tachyon equation in (\ref{mussel}) that
\begin{equation}
T = a_0 \sqrt{y} + \cdots
\end{equation}
where $a_0$ is a constant. Note that the
above solution for the tachyon with its potential included is
indeed the square root of the tachyon solution
given in equation (\ref{whale}) when $V = 0$.
Substituting this value of tachyon in the dilaton equation
in (\ref{mussel}) one obtains the correction
to the restoring force given by
\begin{equation}
- \frac{T e^{\Sigma}}{4 \tau_1^2} = \frac{T}{4 y^2}
(1 - y - a_0^2 y + \cdots)  \; .
\end{equation}
What the above expression implies is that the tachyon back
reaction reduces the restoring force as one moves inside away from
the asymptotically flat region. Thus it indicates that the restoring force
(and hence the tachyon potential) is less and less important as
one moves inside and that it can eventually be neglected,
thus validating our assumption. Furthermore, the potential
$V (T) = \gamma T^2$ itself neglects ${\cal O} (T^3)$ corrections
given, {\em e.g.}\  as in \cite{CST}, by
\begin{equation}
V = \gamma ( T^2 - \frac{T^3}{24} + {\cal O} (T^4) ) \; .
\end{equation}
Thus the correction term
implies that the potential is less steep, suggesting that the potential
energy can eventually be neglected (compared to the kinetic
energy). Also, various
recent works \cite{DL,review,W,M} find instability of the
$d = 2$ string black hole solution when tachyon is included.
Since our solution can be viewed as arising from a similar instability,
it seems that neglecting tachyon potential
is reasonable and that the qualitative features of
our solution will persist even when $V(T)$ is properly
taken into account.

We now show by a perturbative analysis near the horizon when
$V \neq 0$ that the horizon still develops a curvature singularity.
This is seen by an analysis of equations (\ref{betalie}).

We first show that  $x \rightarrow 0 \; (\; \infty)$
in the asymptotically flat region (near horizon).
Recall that $x = - X^2 e^{- \Sigma}$ where $X = \Sigma_1$ and
note that $(\cdots)_1 = \frac{1}{2} (- \gamma)^{\frac{1}{2}}
e^{\Sigma} \frac{d}{d r} (\cdots)$
in $(r, t)$ coordinates described in section 4.
Since the metric in $(r, t)$ coordinates is given by equation
(\ref{rtmetric}), it follows that in the asymptotically flat region
$\Sigma \rightarrow 0$ and also $X \rightarrow 0$. Hence,
$x \rightarrow 0$ there.
Similarly near the horizon $e^{\Sigma} \rightarrow 0$ by
definition and hence $\Sigma \rightarrow - \infty$.
Since $X = \Sigma_1$, it follows that $X \neq 0$ and therefore
$x \rightarrow \infty$ near the horizon. This behaviour
of $x$ tending to zero in the asymptotically flat region and to
infinity near the horizon remains true irrespective
of the nature of tachyon potential and
can be seen explicitly from equations
(\ref{xf1}) and (\ref{xf2}) for special cases
$T = 0$ and $V = 0$ respectively.

Consider the $\beta$-function equations in the form given in
(\ref{betalie}) and (\ref{xfintegral}).
Let us call the last three equations in (\ref{betalie})
as $T, \; f$, and $H$ equation respectively. To find the
solution near $x = 0$ or $\infty$
one first starts with an ansatz for the tachyon field and then
solves $f, \; H$, and $T$ equations, in that order, upto
leading corrections. The equation (\ref{xfintegral}) will be used to
fix constants. The ansatz is considered correct
if the solution of $T$ equation
gives the same form for $T$, to the leading order, as originally assumed.

Consider first $x \rightarrow 0$ and the ansatz
$T = a_0 x^n$ for the tachyon. Let $n > 0$
so that $T$ will not be divergent in the
asymptotically flat region $x \rightarrow 0$. After solving
$f, \; H$, and $T$ equations one gets, for $V = \gamma T^2$,
\begin{eqnarray}
f + f^2 & = & x^{- 1} + f_0^2 - \frac{1}{4} + \cdots  \nonumber \\
H & = & - \frac{a_0}{2} x^{- \frac{1}{4}} + H_0 + \cdots
\nonumber \\
T & = & a_0 x^{\frac{1}{4}} - H_0  \sqrt{x} + \cdots   \; .
\end{eqnarray}
Here and in the following $a_0, \; f_0$, and $H_0$ are constants.
For $V = 0$ one gets $H = H_0 + \cdots$ and
$T = a_0 \sqrt{x} + \cdots$ and $f$ as above.

Next consider the ansatz
$T = a_0 x^n \ln x $ for the tachyon as $x \rightarrow 0$ (with
$n > 0$ for the same reasons as above). After solving
$f, \; H$, and $T$ equations one gets, for $V = \gamma T^2$,
\begin{eqnarray}
f + f^2 & = & x^{- 1} + f_0^2 - \frac{1}{4} + \cdots  \nonumber \\
H & = & - \frac{a_0}{2} x^{- \frac{1}{4}} (\ln x + 4)
+ \cdots   \nonumber \\
T & = & a_0 x^{\frac{1}{4}} \ln x + \cdots   \; .
\end{eqnarray}
The above ansatz is not valid for $V = 0$. In the above
equations $\cdots$
denote the subleading terms in powers of $x$ which can be obtained
by performing more iterations. Note that
for both the ansatzes above, the curvature
scalar $R \; ( = 4 \gamma x f$) tends to zero as $x \rightarrow 0$
as expected and the tachyon behaviour
agrees with that found in \cite{DL}.

We now consider the region near horizon where $x \rightarrow \infty$
and $V \neq 0$. We would like to know
how the solution  behaves near
the horizon, {\em i.e.}\ , $x \rightarrow \infty$, and in particular,
whether the curvature scalar $R$ is divergent or not when
$V \neq 0$. Let us first consider the ansatz
$T = a_0 x^n + \cdots$, where $\cdots$ denote subleading
terms. The exponent $n$ can be positive or negative and $a_0$ is
a constant. After solving
$f, \; H$, and $T$ equations one gets $n \leq \frac{1}{2}$ and
\begin{equation}
T \approx {\rm constant} + \ln x + x^{n - 1}
\end{equation}
schematically, omitting the coefficients.
This is in contradiction to the original ansatz
which, therefore, is incorrect. Next we take the ansatz
$T = a_0 \ln x + \cdots \; $. Solving $f, \; H$, and $T$ equations
we get
\begin{eqnarray}
f + f^2 & = & f_0^2 - \frac{1}{4}
+ \frac{a_0^2}{4} x^{- 1} (\ln^2 x + 2 \ln x + 2) - x^{- 1}
+ \cdots \nonumber \\
H & = & - 2 a_0 f_0 - \frac{a_0}{8 f_0} x^{- 1} (\ln x + 1)
+ \cdots \nonumber \\
T & = & a_0 \ln x + x^{- 1} [ \; \frac{a_0^3}{8 f_0^2}
(\ln^2 x + 4 \ln x + 6)  \nonumber \\
& & - \frac{a_0}{16 f_0^2} (\ln x + 2)
+ \frac{a_0}{2 f_0^2} \; ] + \cdots
\end{eqnarray}
where $\cdots$ denote
${\cal O} (x^{- 2} \ln^* x)$ terms for some integer $*$.
Note that this solution of $T$ agrees to the leading order
with our starting ansatz which, therefore, is correct.
We also get from equation (\ref{xfintegral}) that
\begin{equation}
f_0^2 - \frac{1}{4} = 4 a_0^2 f_0^2
\end{equation}
which implies that $f + f^2$, and hence $f$, is nonzero
as long as $a_0$, and hence, $T$ is nonzero.
Note that the solution given in (\ref{whale}) for $V = 0$
satisfies the above equation, as it should.
One can continue the above iteration to obtain the
subleading terms of ${\cal O} (x^{- 2} \ln^* x)$
but they are not necessary for our purpose.

If the tachyon is not identiaclly zero,
{\em i.e.}\  $a_0 \neq 0$, then as $x \rightarrow \infty, \; f$
approaches a nonzero constant and hence the curvature scalar
given by $R = 4 \gamma x f$ diverges near the horizon where
$x \rightarrow \infty$. Note also that the
leading logarithmic behaviour of $T$
near the horizon when $V \neq 0$
agrees with our solution (\ref{whale}) obtained assuming $V = 0$.

Thus the curvature scalar $R$ is divergent
at the horizon even when the tachyon potential $V (T) \neq 0$.
Hence it follows that when tachyon back reaction
is included, the $d = 2$ string does not admit any static
solution which can be interpreted as a black hole in the usual sense.

\vspace{4ex}

\centerline{\bf 6. Discussions}

\vspace{4ex}

In this section we first briefly present a time dependent
extension of our solutions and then discuss possible methods of
removing the singularities found in previous sections.
Consider the ansatz for the fields $\Sigma, \; \phi$, and $T$
\begin{eqnarray}
\Sigma (\xi, \chi) & = & \Sigma (\xi)
+ \tilde{\Sigma} (\chi)   \nonumber \\
\phi (\xi, \chi) & = & \chi (\xi) + \tilde{\phi} (\chi)   \nonumber \\
T (\xi, \chi) & = & T (\xi) + \tilde{T} (\chi)
\end{eqnarray}
which is the nonlinear analog of seperation of variables.
The variables $\xi$ and $\chi$ are defined in section 2. The equations
for the $\xi$ dependent fields are the same as in (\ref{salmon})
with $e^{\Sigma}$ understood as $e^{\Sigma (\xi, \chi)}$ and those
for the $\chi$ dependent fields are given by
\begin{eqnarray}\label{salxichi}
\Sigma_{0 0} + \phi_0 \Sigma_0  & = & 0 \nonumber \\
\Sigma_{0 0} + \phi_{0 0} + T_0^2  & = & 0 \nonumber \\
\Sigma_1 \phi_0 + \Sigma_0 \phi_1 - 2 T_0 T_1 & = & 0 \nonumber \\
T_{0 0} + \phi_0 T_0  & = & 0 \nonumber \\
\phi_{0 0} + \phi_0^2 & = & 0
\end{eqnarray}
where we have omitted the tildes over the $\chi$ dependent fields.
As before $(\cdots)_0 \equiv (\chi \frac{d}{d \chi}) (\cdots)$
and $(\cdots)_1 \equiv (\xi \frac{d}{d \xi}) (\cdots)$. Note that
the above equations are the same as those given in (\ref{salmon}) with
$\gamma = 0$ and $\xi$ replaced by $\chi$.
{}From the first, fourth, and fifth equations above it immediately
follows that
\begin{equation}
\Sigma_0 = a_1 \phi_0, \; \; \;
T_0 = a_2 \phi_0
\end{equation}
where $a_1$ and $a_2$ are constants. The second equation in
(\ref{salxichi}) gives
\begin{equation}
a_2^2 = 1 + a_1
\end{equation}
while the third equation implies, using (\ref{whale}) for
$\Sigma(\xi), \; \phi(\xi)$, and $T(\xi)$, that
\begin{equation}
a_1 = 0, \; \; a_2^2 = (4 \epsilon (1 + \epsilon))^{- 1} \; .
\end{equation}
The splitting of the $\beta$-function equations (\ref{betaxichi})
into (\ref{salmon}) and (\ref{salxichi}) and the
use of the solutions (\ref{whale}) of equations (\ref{salmon}),
which involve $e^{\Sigma}$ terms, is justified {\em a posteriori}
since $a_1 = 0$ and hence $\Sigma = \Sigma (\xi)$.
The equation for $\phi (\chi)$ can be
easily solved and we get
\begin{eqnarray}
\phi (\chi) & = & \ln (b_1 \ln A \chi) \nonumber \\
T (\chi) & = & \ln (b_2 \ln A \chi)
\end{eqnarray}
where $A, \; b_1$, and $b_2$ are constants. Note however that this
time dependent extension does not alter any of the features of
the previous static solution (\ref{whale}). Nor is it likely to
represent the general solution of the $\beta$-function equations
(\ref{betaxichi}).

We now discuss some
possible interaction terms that may remove the singularities found
in previous sections, atleast the new ones:

(1) Higher order $\alpha'$ corrections: Tseytlin \cite{tseyt} had shown
that black hole solutions of \cite{W,M, rest} survive these
corrections. Very likely, the solution given here will also survive
these corrections since the tachyon can be thought of as
an (anti)damped
oscillator gaining energy by gravitational interactions --- so that
it would have grown strong before one reaches
the region of strong curvature
where $\alpha'$ corrections are deemed important.
Thus these corrections may not remove the singularities.

(2)Higher massive modes :
For similar reasons as above
higher massive modes $A_n$ are of no help. In fact,
taking their effective action as given in \cite{CST} with zero potential,
it is easily seen that $A_n$'s have solutions similar to that of
tachyon $T$ and hence do not remove the singularities.

(3) Antisymmetric tensor, $H$ (indices on $H$ suppressed) :
This field is not there for $d = 2$ space-time. However,
for $d = 2$ toy models of a $D$ dimensional space-time, as
considered in \cite{CGHS,review} for example,
the resulting equations that include quantum effects
will be similar to the $\beta$-function equations for $d = 2$ strings.
$H$ field interactions present in such cases may possibly
remove the singularities.
Moreover, $H$ fields invariably arise in any space-time
obtained from string theory, so it is natural
to include them.

(4) Supersymmetry : This symmetry introduces
fermions which may provide enough
repulsive force to avoid the formation of the singularities.
On the other hand, the fermions might instead form attractive condensates
and not remove the singularities.

(5) Other approaches :
It will also be interesting to find an analog of our solution,
which includes tachyon back reaction, in the context of
gauged Wess-Zumino-Witten models such as considered in \cite{W}
or in the context of matrix model where tachyon equation
in the black hole background has been found recently
\cite{blackmatrix}. Perhaps in these contexts one may get a
better insight into the singularities and methods of avoiding them.

The above (and other) possibilities are
worth pursuing. It is important to resolve this problem of new
singularities --- to understand them better and to remove them
if possible. This issue is particularly relevent
for the string inspired toy models of $d = 2$ quantum gravity
that may answer the puzzles of $d = 4$ space-time.
The removal of the singularities seen here might
also suggest new interactions that could be important.

\vspace{3ex}

I thank S. Sen for encouragement and S. R. Das for
pointing out the last paper in \cite{DL}.
This work is supported by EOLAS Scientific Research Program
SC/92/206.

{\bf Notes Added:}

(1) N. E. Mavromatos has kindly brought ref. \cite{mavro}
to our attention where time dependent perturbations of the black
hole solution are considered perturbatively
in the asymptotically flat region.

{\bf Notes Added in Proof:}

(1) In ref. \cite{rbmann} Mann et al. consider $1 + 1$ dimensional
black hole formation in various models, for the first time
to our knowledge by two sided gravitational collapse.
For the string inspired models they find
that black hole formation requires dilaton
surface charges, interpreted as arising from a tachyon potential.

\vspace{3ex}

We comment on two recent papers \cite{marcus,peet} which appeared
after our work:

\vspace{3ex}

(2) In ref. \cite{marcus} Marcus and Oz
calculate tachyon effects in $d = 2$ string black hole
and find none of the features presented here. This is not surprising
since they work with the
tachyon equation in black hole background, thus
not taking into account the effects of backreacted graviton and dilaton
on tachyon. Therefore tachyon back reaction is not
completely accounted for in their work.

\vspace{3ex}

(3) In ref. \cite{peet} Peet et al. point out that besides the
singular solution described in this paper, the $d = 2$ string
admits another solution which is regular at the horizon. It can be
obtained by taking the ansatz $T = const. \, + \, \ldots$ for
the tachyon near the horizon in equations (16) and (17). This solution
is trivial (that is, $T = const.$) when the tachyon potential
$V = 0$ and has infinite energy when $V \neq 0$ (see \cite{peet}).
These authors further argue that a singular configuration cannot
be dynamically formed starting from a regular one. We find their
argument inconclusive and
will address this issue elsewhere \cite{future}.

\end{document}